\documentstyle[11pt,epsfig]{article}
\title{\bf Weyl-Dirac theory predictions on galactic scales}
\author{S. Mirabotalebi$^{1}$\thanks{e-mail:~s$_{-}$mirabotalebi@iust.ac.ir.},
S. Jalalzadeh$^{2}$\thanks{e-mail:~s-jalalzadeh@sbu.ac.ir.}, M.
Sadegh Movahed$^{2}$\thanks{e-mail:~m.s.movahed@ipm.ir.} and
H. R. Sepangi$^{2}$\thanks{e-mail:~hr-sepangi@sbu.ac.ir} \vspace{0.5cm}\\
$^1${\small Department of Physics, Iran University of Science and
Technology
 (IUST), Narmak, Tehran, Iran}\\
$^2${\small Department of Physics, Shahid Beheshti University,
Evin, Tehran 19839, Iran.}\\ }
\begin{document}
\maketitle
\begin{abstract}
We consider the Weyl-Dirac theory within the framework of the weak
field approximation and show that the resulting gravitational
potential differs from that of Newtonian by a repulsive correction
term increasing with distance. The scale of the correction term
appears to be determined by the time variation rate of the
gravitational coupling. It is shown that if the time variation rate
of gravitational coupling is adopted from observational bounds, the
theory can explain  the rotation curves of typical spiral galaxies
without resorting to dark matter. To check the consistency of our
theoretical model with observation we use Likelihood analysis to
find the best-fit values for the free parameters. The mean value for
the most important free parameter, $\beta\times 10^{14} (1/yr)$,
using the Top-Hat and Gaussian priors are
$6.38^{+2.44}_{-3.46}\!_{-6.71}^{+6.18}$ and
$5.72_{-1.18}^{+1.22}\!_{-2.69}^{+2.90}$, respectively. Although the
interval for which $\beta$ is defined is wide, our results show that
the goodness of the fit is, by and large, not sensitive to this
quantity. The intergalactic effects and gravitational lensing of
clusters of galaxies are estimated and seem to be consistent with
observational data.
\end{abstract}
\section{Introduction}

It has long been known that standard gravitational theories cannot
correctly predict the dynamics of large astronomical systems. The
rotation curves of spiral galaxies and gravitational lensing  are
among the well known examples \cite{nar,bn}. The rotation
velocities usually tend to a constant or slightly rising values as
a function of distance from the center of galaxy. This is in sharp
contrast to the inverse square force law which implies a decline
in velocity. The gravitational lensing of clusters of galaxies is
the other observation that appears to be in conflict with the
standard gravitational theories on extra galactic scales
\cite{bn}. These observations are usually explained by postulating
the existence of dark matter \cite{ash,bn}. Since there is no
established observation of dark matter, the belief that standard
gravitational theories should be modified on large length scales
has been gaining momentum in recent years. Many attempts have
emerged in this regard recently \cite{bekmil}-\cite{mof}.

The purpose of the present paper is to undertake an investigation
on the predictions of the dynamics of large astronomical systems
in the Weyl-Dirac theory \cite{we} which, in effect, is the Weyl
theory \cite{we} modified by Dirac later on. Dirac made use of the
theory to provide a framework to explain his large number
hypothesis. This modified theory can be considered as a
scalar-tensor theory of gravity and leads to a time varying
gravitational coupling $G$ in accordance with the Dirac
hypothesis. In comparison with Weyl theory, this modified theory
is simpler and in agrement with the general theory of relativity.
As we shall show below, its implications on the dynamics of
astronomical objects is significant and therefore well worth
studying.

The organization of this paper is as follows: in the next section we
briefly discuss the ingredients of the Weyl geometry followed by the
introduction of the Weyl-Dirac model in section 3. In section 4 we
present the spherically symmetric solutions of the model for a
point-like mass source \cite{jamo}. In section 5 we generalize the
solutions to galaxies considered as spherically symmetric extended
mass sources. The consistency of our model with observational
rotation curves of typical spiral galaxies using Likelihood
statistics will be checked and the best-fit values for free
parameters of the model are determined in section 6. Sections 7 and
8 are devoted to the intergalactic effects and the gravitational
lensing. Concluding remarks are presented in section 9.

In what follows we shall use units in which $\hbar=c=G=1$.
However, in order to interpret physical results we turn to
conventional units from time to time. The signature of the metric
is taken as (-,+,+,+) and the convention for the sign of the
curvature tensor is that of Misner, Thorne and Wheeler \cite{mtw}.
For simplicity, the scalar field indices are taken as their
covariant derivatives.

\section{Weyl geometry}

Weyl geometry is a natural generalization of the Riemannian
geometry \cite{we}. In Weyl geometry the length of a vector is
assumed to change as well as its direction in a parallel
displacement. This means that if a vector has length $l$ at a
point with coordinates $x^{\mu}$, then
\begin{equation}\label{I1}
\delta l=(k_{\mu}\delta x^{\mu})l\,,
\end{equation}
after a parallel displacement $\delta x^{\mu}$. The covariant
vectors $k_{\mu}$ are considered as the field quantities and may
be called the Weyl meson fields. By equation (\ref{I1}), the total
displacement around a small closed loop is

\begin{equation}\label{I2}
\delta l =l F_{\mu\nu}\delta S^{\mu\nu}\,,
\end{equation}
where

\begin{equation}\label{I3}
F_{\mu\nu}=k_{\nu,\mu}-k_{\mu,\nu}\,,
\end{equation}
is called length curvature \cite{per} and $\delta S^{\mu\nu}$ is
the element of the area enclosed by the loop. It follows from
definition (\ref{I1}) that the comparison of elements of length at
two different points which are not separated by an infinitesimal
distance can be made only with respect to a path joining the
points and obviously, different paths lead to different results.
The comparison then is possible if one arbitrarily defines the
standards of length at each space-time point. Alternatively, the
change in length $l$ can be parameterized by means of an arbitrary
function $\Omega(x_{\mu})$ such that

\begin{equation}\label{I4}
l'=\Omega(x_{\mu})l\,,
\end{equation}
and
\begin{equation}\label{I5}
k'_{\mu}=k_{\mu}+(\ln[\Omega(x_{\mu})])_{,\mu}\,.
\end{equation}
Such a transformation does not change the tensor field
$F_{\mu\nu}$ defined in equation (\ref{I3}). One should note that
the tensor field $F_{\mu\nu}$ satisfies
\begin{equation}\label{I6}
\nabla _{\tau}
F_{\mu\nu}+\nabla_{\nu}F_{\tau\mu}+\nabla_{\mu}F_{\nu\tau}=0\,.
\end{equation}
Equations (\ref{I3}) and (\ref{I6}) are  similar to Maxwell
equations relating the vector potential to the electromagnetic
tensor. It is therefore natural to interpret $k_{\mu}$ and
$F_{\mu\nu}$ in just the same manner and transformations
(\ref{I5}) as gauge transformations \cite{abs}. Hence the meson
fields $k_{\mu}$ may be recognized as photons. However on the
basis of the axiomatic formulation of spacetime theory presented
by Ehlers, Pirani and Schild \cite{eps}, this interpretation is
unacceptable. It seems that the length curvature $F_{\mu\nu}$ must
be related to a phenomena which has not been observed in nature.
It has been shown that Weyl mesons do not interact with leptons or
quarks and other vector mesons in minimal form
\cite{hk}-\cite{ch2}. They only interact with gravitons and scalar
mesons, that is, the  Higgs fields. The question then naturally
arises as to whether Weyl mesons may account for at least part of
the dark matter in our universe. Some attempts on this subject can
be found which consider the global structure of the universe
\cite{I1}-\cite{I4}. In this paper we concentrate on the
possibility of the Weyl-Dirac theory predicting dark matter
effects on the dynamics of astronomical objects.

\section{The model}
Consider the Weyl-Dirac action functional \cite{di}

\begin{equation}\label{c1}
S[ \phi , k_{\mu} ] = \frac{1}{2}\int d^{4}x \sqrt{-g}\left\{
\frac{1}{2} F_{\mu\nu}F^{\mu\nu} +\, ^{*}R\phi^{2} + \alpha
\phi_{*\mu}\phi^{*\mu} \right\}\,,
\end{equation}
where $\alpha$ is a dimensionless constant and $\phi^{*}$ is the
co-covariant derivative of the scalar field $\phi$ and is defined
as\footnote{In Weyl's geometry, a localized quantity $\psi$ is
called a co-tensor if under local unit transformation, it
transforms as $\psi^{'}=\Omega^{n}(x_{\mu})\psi$. Then $\psi$ is
said to be of power $n$. Also, a co-covariant derivative is a
modified covariant derivative and is a co-tensor. For a scalar
field $S$ it is defined as

\begin{equation}\label{c2}
S_{*\mu}=S_{\mu}-nk_{\mu}S\,.
\end{equation}
Here $\phi$ is a scalar field of power $n=-1$.}

\begin{equation}\label{c3}
\phi_{*\mu}=\phi_{\mu}+ k_{\mu}\phi \,,
\end{equation}
where $^{*}R$ is the modified curvature scalar given by
\begin{equation}\label{c4}
^{*}R=R+6k_{;\mu}^{\mu}-6k_{\mu}k^{\mu}\,,
\end{equation}
with $R$ being the scalar curvature. The original Weyl action
contains the term $(^{*}R)^{2}$ instead of the term $^{*}R
\phi^{2}$. Dirac proposed the term $^{*}R\phi^{2}$, thereby
avoiding  the great complication of the Weyl action. Using
definitions (\ref{c3}) and (\ref{c4}) one arrives at
\begin{eqnarray}
\label{c5}S[\phi,k_{\mu}]&=&\frac{1}{2}\int d^{4}x
\sqrt{-g}\left\{\frac{1}{2}F_{\mu\nu}F^{\mu\nu}
+R\phi^{2}+\alpha\phi_{\mu}\phi^{\mu}
+(\alpha-6)\phi^{2}k_{\mu}k^{\mu}+ 2(\alpha-6)\phi
k^{\mu}\phi_{\mu}\right.\nonumber\\
& &\left. +6(\phi^{2}k^{\mu})_{;\mu}\right\},
\end{eqnarray}
where the term $6(\phi^{2}k^{\mu})_{;\mu}\sqrt{-g}$ is a total
differential and can be ignored.

Action (\ref{c5}) is invariant under local transformations of
units, that is under general conformal transformations. Here,
local unit transformations are taken to be composed of gauge
transformations defined in equations (\ref{I2}) and (\ref{I3})
with
\begin{equation} g_{\mu\nu}\longrightarrow
\Omega^{2}(x_{\mu})g_{\mu\nu}\,,\label{c6}\end{equation}
\begin{equation}
\phi\longrightarrow \Omega^{-1}(x_{\mu}) \phi\,
\label{m42}\,.
\end{equation}
In order to incorporate the matter fields, we add the
corresponding action $S_{m}$ to (\ref{c5}) and consider $S_{m}$ as
being built out of matter fields in the usual manner. Variation of
$S[\phi,k_{\mu}]+S_{m}$ with respect to $g_{\mu\nu}$, $\phi$ and
$k_{\mu}$ respectively gives
\begin{equation}
G_{\mu\nu}=\phi^{-2} \left\{T_{\mu\nu} + E_{\mu\nu}+
\Theta_{\mu\nu}+\Sigma_{\mu\nu}\right\}\,,
\label{c7}\end{equation}
\begin{equation}
\alpha\phi^{\mu}_{;\mu} - R \phi -(\alpha-6)\left\{\phi
k_{\mu}k^{\mu}- \phi k^{\mu}_{;\mu}\right\}-\psi= 0\,,
\label{c8}\end{equation}
\begin{equation}\label{c9}
-(F^{\nu\mu})_{;\mu}+(\alpha-6)(\phi^{2}k^{\nu}+\phi\phi^{\nu})+J^{\nu}=0\,.
\end{equation}\\
Here $G_{\mu\nu}= R_{\mu\nu}-\frac{1}{2}g_{\mu\nu}R$ is the
Einstein tensor and
\begin{equation}\label{c10}
T_{\mu\nu}=\left(-2+\frac{\alpha}{2}\right)
g_{\mu\nu}\phi_{\alpha}\phi^{\alpha}+(2-\alpha)
\phi_{\mu}\phi_{\nu}-2g_{\mu\nu}\phi(\phi^{\alpha})_{;\alpha}+
2\phi\phi_{\mu;\nu}\,,\end{equation}
\begin{equation}\label{c11}
E_{\mu\nu}=\frac{1}{4}F_{\alpha\beta}F^{\alpha\beta}g_{\mu\nu}-
F_{\mu\alpha}F_{\nu\beta}g^{\alpha\beta}\,,
\end{equation}
\begin{equation}
\Theta_{\mu\nu}= (\alpha-6)\left\{ \phi^{2}\left(- k_{\mu}k_{\nu}+
\frac{1}{2}g_{\mu\nu} k^{\alpha} k_{\alpha}\right)-\phi\left(
k_{\mu}\phi_{\nu}+ k_{\nu} \phi_{\mu}-
k_{\alpha}\phi^{\alpha}g_{\mu \nu}\right)\right\}\,
,\label{c12}\end{equation} where $\Sigma_{\mu\nu}$ is the stress
tensor of the matter fields defined as

\begin{equation}
\Sigma_{\mu\nu}= - \frac{2}{\sqrt{-g}}\frac{\delta S_{m}} {\delta
g^{\mu\nu}}\,,\label{c13}\end{equation} and $J^{\mu}$ denotes the
current density vector

\begin{equation}\label{c14}
J^{\mu}=\frac{\delta S_{m}}{\delta k_{\mu}}\,,
\end{equation}
with  $\psi$ given by

\begin{equation}\label{c}
\psi=\frac{\delta S_{m}}{\delta \phi}\,.
\end{equation}
It should be noted that equations (\ref{c7})-(\ref{c9}) are not
independent.

Now, taking the trace of equation (\ref{c7}), using definitions
(\ref{c10})-(\ref{c12}) and comparing the result with equations
(\ref{c8}) and (\ref{c9}), one finds
\begin{equation}\label{c17}
\Sigma_{\mu}^{\mu}+(\alpha-6)\left[\phi^{2}k^{\mu}+\phi\phi^{\mu}\right]_{;\mu}+\phi\psi=0\,,
\end{equation}
where from (\ref{c9}) we have
\begin{equation}\label{c16}
(\alpha-6)\left[\phi^{2}k^{\mu}+\phi\phi^{\mu}\right]_{;\mu}+
J^{\mu}_{;\mu}=0\,.
\end{equation}
Taking the divergence of (\ref{c7}) and using (\ref{c16}) leads to

\begin{equation}
\Sigma^{\mu\nu}_{;\mu}-\frac{\phi_{\nu}}{\phi}\Sigma^{\mu}_{\mu}=
J^{\mu}F_{\mu}^{\nu}-\left(k^{\nu}+\frac{\phi_{\nu}}{\phi}\right)
J^{\mu}_{;\mu}\,. \label{c15}\end{equation} One may assign to the
non-gravitational part of the fields a stress tensor defined as

\begin{equation}\label{c18}
\tau_{\mu\nu}=T_{\mu\nu}+E_{\mu\nu}+\Theta_{\mu\nu}\,.
\end{equation}
It may be proved \cite{r}, using equation (\ref{c16}) that
\begin{equation}\label{c19}
\tau^{\mu\nu}_{;\mu}-\frac{\phi_{\nu}}{\phi}\tau^{\mu}_{\mu}=
J^{\mu}F_{\mu}^{\nu}-\left(k^{\nu}+\frac{\phi_{\nu}}{\phi}\right)
J^{\mu}_{;\mu}\,.
\end{equation}
From equations (\ref{c15}) and (\ref{c19}) we find the following
relation for the total stress tensor

\begin{equation}\label{c20}
(\Sigma^{\mu\nu}+\tau^{\mu\nu})_{;\mu}-\frac{\phi_{\nu}}{\phi}(\Sigma^{\mu}_{\mu}+\tau^{\mu}_{\mu})=0\,.
\end{equation}
It is worth noting that equation (\ref{c15}) may be used to derive
the equation of motion for a test particle \cite{I5,r}.

To progress further, let us take the matter content in our model
as having consisted of identical particles with rest mass $m$ and
the Weyl charge $q$ in the form of dust, so that

\begin{equation}\label{c21}
\Sigma^{\mu\nu}=\rho u^{\mu}u^{\nu}\,,
\end{equation}
where $u^{\mu}$ is the 4-velocity and the mass density $\rho$ is
given by

\begin{equation}\label{c22}
\rho=m \rho_{n}\,,
\end{equation}
with $\rho_{n}$ being the particle density. Making use of the
conservation of the number of particles, we find using equation
(\ref{c15}), the equation of motion

\begin{equation}\label{c23}
\frac{du^{\mu}}{ds}+\left\{\begin{array}{c}
\mu\\
\nu \lambda\\
\end{array}\right\}u^{\nu}u^{\lambda}-\frac{\phi_{\nu}}
{\phi}(g^{\mu\nu}-u^{\mu}u^{\nu})= \frac{q}{m}u_{\nu}F^{\nu\mu}\,,
\end{equation}
where $\left\{\begin{array}{c}
\mu\\
\nu \lambda\\
\end{array}\right\}$ denotes the Riemann  Christoffel symbol. In the left hand side of equation (\ref{c23}) the term
containing $\frac{\phi_{\nu}} {\phi}$ may be taken as a variable
mass term for the particle. In such a case we have \cite{r}
\begin{equation}\label{c25}
m=m_{0}\phi\,,
\end{equation}
where $m_{0}$ is a constant.

\section{Field of a mass source}

Let us study the field of a mass source at rest at the origin and
assume that it is in a current-free region. In such a case we can
consider the vacuum solutions of the field equations, that is we
take $\Sigma_{\mu\nu}=0$ and $J_{\mu}=0$. Now consider the
spherically symmetric line-element

\begin{equation}\label{s1}
ds^{2}=-e^{\upsilon}dt^{2}+e^{\lambda}dr^{2}+r^{2}d\Omega^{2}\,,
\end{equation}
where $\upsilon$ and $\lambda$ are functions of $t$ and $r$.
Neglecting the factor $\alpha-6$  which can be shown to be small
\cite{r,m} in equation (\ref{c9}), we find, using the above metric

\begin{equation}\label{s2}
k_{0,r}= \frac{\gamma(t)e^{(\upsilon+\lambda)}}{r^{2}},
\end{equation}
where $\gamma(t)$ is an arbitrary function of time $t$ resulting
from the integration. Using relation (\ref{s2}) in equation
(\ref{c7}), one finds

\begin{equation}\label{s3}
e^{-\lambda}\left(
-\frac{\lambda_{r}}{r}+\frac{1}{r^{2}}\right)-\frac{1}{r^{2}}=\frac{1}{\phi}e^{-\upsilon}\left(\phi_{t}
\lambda_{t}+\frac{3\phi_{t}^{2}}{\phi^{2}}\right)-\frac{\gamma^{2}}{r^{4}\phi^{2}},
\end{equation}

\begin{equation}\label{s4}
e^{-\lambda}\left(
-\frac{\upsilon_{r}}{r}+\frac{1}{r^{2}}\right)-\frac{1}{r^{2}}=\frac{1}{\phi}e^{-\upsilon}\left(\phi_{t,t}
-\frac{\phi_{t}^{2}}{\phi^{2}}-\phi_{t}\upsilon_{t}\right)-\frac{\gamma^{2}}{r^{4}\phi^{2}},
\end{equation}

\begin{equation}\label{s5}
-\frac{\lambda_{t}}{r}=\frac{\phi_{t}\upsilon_{t}}{\phi}.
\end{equation}
The form of the left hand side of equations (\ref{s4}) and
(\ref{s5}) suggest that

\begin{equation}\label{s6}
e^{\upsilon} = f(t)e^{-\lambda}.
\end{equation}
Using equations (\ref{s5}) and (\ref{s6}) we see that equations
(\ref{s3}) and (\ref{s4}) are identical provided $f(t)$ satisfies

\begin{equation}\label{s7}
\frac{f_{t}}{f}=2\frac{\phi_{t,t}}{\phi_{t}}-4\frac{\phi_{t}}{\phi}.
\end{equation}
The last expression gives

\begin{equation}\label{s8}
f=\frac{\phi_{t}^{2}}{\beta^{2}\phi^{4}},
\end{equation}
where $\beta$ is a constant parameter. We are now in a position to
obtain the generalization of the results derived in \cite{r}, that
is

\begin{equation}\label{s9}
e^{\upsilon}=\frac{\phi_{t}^{2}}{\beta^{2}\phi^{4}}e^{-\lambda},
\end{equation}
with

\begin{equation}\label{s10}
e^{-\lambda}=\frac{1}{2}-\frac{m}{\phi
r}+\frac{q^{2}}{\phi^{2}r^{2}}+\Delta,
\end{equation}

\begin{equation}\label{s11}
\Delta=\left[\left(\frac{1}{2}-\frac{m}{\phi
r}+\frac{q^{2}}{\phi^{2}r^{2}}\right)^{2}+\beta^{2}\phi^{2}r^{2}\right]^{\frac{1}{2}},
\end{equation}
where we have assumed $\gamma(t)=\frac{q}{\phi}$. The parameters
$m$ and $q$ are constant and may be interpreted as mass and charge
of the source.

The above solutions are valid for any gauge function $\phi(t)$. We
choose to work in the Einstein gauge by taking ($ds^{2}\rightarrow
d\bar{s}^2=\phi^{2}ds^{2}, \phi\rightarrow\bar{\phi}=1$). In this
case one gets

\begin{equation}\label{s12}
d\bar{s}^{2}=-e^{-\lambda}dT^{2}+e^{2\beta
T}\left(e^{\lambda}dr^{2}+r^{2}d\Omega^{2}\right),
\end{equation}
where

\begin{equation}\label{s13}
dT=\frac{\phi_{t}^{2}}{\beta\phi}dt=\frac{d\phi}{\beta \phi},
\end{equation}
which immediately yields
\begin{equation}\label{s14}
\phi=\phi_{0}e^{\beta T}.
\end{equation}
It should be noticed that the factor $\phi^{-2}$ associated with
$\Sigma_{\mu\nu}$ in (\ref{c7}) can be considered as the
gravitational coupling $G$ in natural units ($\hbar=c=1$). Hence
we may consider

\begin{equation}\label{s15}
G\equiv \phi^{-2}.
\end{equation}
Using (\ref{s14}), the last expression leads to
\begin{equation}\label{s16}
G=G_{0} e^{-2\beta T},
\end{equation}
where $G_{0}=\phi_{0}^{-2}$ may be interpreted as the Newtonian
coupling constant. Therefore, the exponential factor $e^{-2\beta
T}$ shows the time evolution of $G$ where $T$ is the elapsed time.
From (\ref{s16}) we find

\begin{equation}\label{s17}
\frac{G_{T}}{G}=-2\beta.
\end{equation}
The empirical measurements on the time evolution of $G$ shows that
\cite{wtbt}

\begin{equation}\label{s18}
\frac{G_{T}}{G}=(4 \pm 9)\times 10^{-13} \, \mbox{yr}^{-1}.
\end{equation}
This sets the following restriction on $\beta$
\begin{equation}\label{s19}
-6.5\times10^{-13} \leq\beta\leq 2.5\times10^{-13}\,
\mbox{yr}^{-1}.
\end{equation}

\section{Rotational velocity of galaxies}

In this section we apply the results obtained above to the
rotational velocity of typical spiral galaxies. In order to obtain
the gravitational potential $U$ of the point mass source $m$
considered in section 4, we first note from (\ref{c25}) that in
the Einstein gauge we have $m=m_{0}$. Now let us consider the weak
field limit

\begin{equation}\label{r0}
g_{00}\cong -(1+2U).
\end{equation}
From (\ref{s12}) and (\ref{s10}) in the conventional units, this
leads to

\begin{equation}\label{r1}
U\approx-\frac{m_{0}G}{r}+4
\left(\frac{m_{0}G}{c^{2}}\right)^{3}\frac{\beta^{2}}{r}+2\left(\frac{m_{0}G}{c^{2}}\right)^{2}\beta^{2}
+\left(\frac{m_{0}G}{c^{2}}\right)\beta^{2}r
+\frac{\beta^{2}r^{2}}{2}\,.
\end{equation}
where we have neglected the term $\frac{1}{r^{2}}$ for large $r$.
As it is clearly seen, equation (\ref{r1}) differs from the
Newtonian gravitational potential by a number of correction terms.
The presence of the variable gravitational coupling $G$ may
equivalently account for this discrepancy, in accordance with
reference \cite{sid}. One should note that parameter $\beta$
appears in all the correction terms. Taking into account the
possible bound on $\beta$ which is given in (\ref{s19}), the
correction terms are too small even if we consider $m_{0}$ to be
of the order of a typical galactic mass. However, the last term,
$\frac{\beta^{2}r^{2}}{2}$, may result in significant effects at
galactic scales. Therefore, the first and second terms of
(\ref{r1}) may have important effects on galactic and larger
scales. Hence we consider the gravitational potential of a point
mass source as follows
\begin{equation}\label{r2}
 U(r)\approx -\frac{m_{0}G}{r}+\frac{\beta^{2}r^{2}}{2}.
\end{equation}
This gravitational potential differs from the Newtonian one by a
repulsive correction term.

In order to study the consequences of the gravitational potential
(\ref{r2}) on galactic objects let us generalize the gravitational
potential $U$ for an extended object taken as a galaxy. To this
end one may treat the exterior of a galaxy as a vacuum solution
and its interior as filled with matter and discuss the junction
conditions along the surface where the two solutions are to be
joined. Alternatively, one may treat the whole space as filled
with matter with an appropriate fall off condition on the density.
In fact, we adopt the second strategy in a weak field
gravitational approximation. This strategy has also been
considered in \cite{m, mk, ms, kkos, mof}. We assume that a galaxy
contains galactic objects such as solar systems which, on galactic
scales, would be seen as point like sources and the gravitational
potential for each individual galactic object obeys the form $U$
given by equation (\ref{r2}) when $m_{0}=M_{\odot}$, where
$M_{\odot}$ denotes the solar mass. We shall consider the matter
density of these objects satisfying an appropriate fall off
condition on galactic scales. In a weak gravitational field
approximation the gravitational potential of the galaxy as a
function of $r$, that is, the distance from the center of the
galaxy, is given by

\begin{equation}\label{r}
V(r)= \int \frac{4\pi\rho_{M}(r')}{M_{\odot}}U(|r-r'|)r'\,^{2} d
r'.
\end{equation}
Here $\rho_{M}(r')$ is the mass density of the galaxy. We shall
consider a galaxy core model for the mass distribution by assuming
a spherically symmetric galaxy with a core density $\rho_{c}$
within a core radius $r < r_{c}$. It should be noted that in order
to  more accurately describe the core behavior, one may consider
the bulge or thin disc effects according to the observational
distribution of the luminous matter of a typical galaxy. However,
the model we consider here yields a reasonable description of the
velocity rotation curves without taking the above mentioned
effects into account.

From (\ref{r}) the gravitational potential can be obtained as
follows

\begin{equation}\label{r3}
V(r)= -\frac{G {\cal M}(r)}{r}+\frac{\beta^{2}M}{2M_{\odot}}r^{2}+
\frac{\beta^{2}I_{cm}}{2M_{\odot}}.
\end{equation}
In equation (\ref{r3}), M is the galaxy mass

\begin{equation}\label{}
M=M_{*}+M_{HI}+M_{DB},
\end{equation}
with $M_{*}$, $M_{HI}$ and $M_{DB}$ respectively denoting the
visible mass, the neutral hydrogen mass and the possible dark
baryon mass and gas with $I_{cm}$ being the moment of inertia
about the center of mass and

\begin{equation}\label{r4}
{\cal M}(r)= 4\pi \int_{0}^{r} dr'\, r'\,^{2} \rho_{c}(r'),
\end{equation}
is the mass inside the luminous core of the galaxy described by a
ball of radius $r=r_{c}$. Inside the ball the dynamics is
described by Newtonian theory and the effect of the last two terms
in (\ref{r3}) is negligible. Moreover outside the core the effect
of the second term becomes more important. A simple model of
${\cal M}(r)$ is given by \cite{mof}

\begin{equation}\label{r5}
{\cal
M}(r)=M\left(\frac{r}{r+r_{c}}\right)^{3\gamma},\end{equation}
where
\begin{equation}\label{r7}
\gamma=\left\{\begin{tabular}{cc}
1 & HSB  galaxies  \\
2 & LSB  and  dwarf galaxies. \\
\end{tabular}\right.
\end{equation}
Well outside the core radius, namely for $r\gg r_{c}$, from
(\ref{r5}) we have
\begin{equation}\label{r8}
{\cal M}(r) \longrightarrow M.
\end{equation}
Also, well inside the core radius the density $\rho(r)$ has a
constant value for HSB galaxies and $\rho(r)\propto r^{3}$ for LSB
and dwarf galaxies. From (\ref{r3}) the square of the rotational
velocity of a solar object about the center of a galaxy can be
obtained as
\begin{equation}\label{r6}
v^{2}(r)\approx \frac{{\cal M}(r)G}{r} +
\frac{M}{M_{\odot}}\beta^{2}r^{2}\,.
\end{equation}

\section{Observational constraint on the model parameters}

To find the consistency of this model with observational data, we
compare the predictions of our model with data directly obtained
from observations and find the best fitting parameters. Since one
cannot expect the theory to exactly explain observational data, we
give confidence intervals for the free parameters of the model
using likelihood analysis.

To begin with, using equation (\ref{r5}), we rewrite equation
(\ref{r6}) as follows
\begin{equation}\label{r67}
v^{2}(r)=\frac{\alpha}{r}\left(\frac{r}{r+r_c}\right)^{3\gamma}+\alpha\beta^{2}r^{2}\,.
\end{equation}
where $\alpha=M/M_{\odot}$. In order to compare the theoretical
results with the observational data, we must compute the
rotational velocities given by equation (\ref{r67}). For this
purpose, we redefine free parameters of the model as
\begin{eqnarray}
\alpha&=&10^{10}\times\bar{\alpha}\\
\beta&=&10^{-14}\times \bar{\beta}\quad \left(\frac{1}{yr}\right)
\end{eqnarray}

Let us compute the quality of the fitting through the least
squared fitting quantity $\chi^2$ defined by
\begin{eqnarray}\label{chi}
\chi^2(\bar{\alpha},\bar{\beta},r_c)&=&\sum_{i}\frac{[v_{{\rm
obs}}(r_i)-v_{{\rm
th}}(r_i;\bar{\alpha},\bar{\beta},r_c)]^2}{\sigma_i^2},
\end{eqnarray}
where $\sigma_i$ is the observational uncertainty in the
rotational velocity. To constrain the parameters of model, we use
the Likelihood statistical analysis
\begin{eqnarray}
{\cal
L}(\bar{\alpha},\bar{\beta},r_c)={\mathcal{N}}e^{-\chi^2(\bar{\alpha},\bar{\beta},r_c)/2},
\end{eqnarray}
where ${\mathcal{N}}$ is a normalization factor. In the presence
of every nuisance parameters, Likelihood function should be
marginalized (integrated out). Using equation (\ref{chi}) we can
find the best fit-values of the model parameters as the values
that minimize $\chi^2(\bar{\alpha},\bar{\beta},r_c)$. Table I
shows different priors on the model parameters used in the
likelihood analysis.

\begin{center}
\textbf{Table I:  Priors on the parameter space, used in the
likelihood analysis.}\vspace{2mm}\\
\begin{tabular}{|c|c|c|}
\hline {\rm Parameters}  & {\rm prior}& \\ \hline&&\\
$\bar{\alpha}$&$[0-15]$ & {\rm Top-Hat}
\\&&\\\hline&&\\
$\bar{\beta}$ &$[0-20]$&{\rm Top-Hat}\\&&\\\hline
 &&\\$\bar{\beta}$
&$10.0\pm0.5$&{\rm Gaussian}\\&&\\\hline
 &&\\$r_c$ &$[0-7]{\rm kPc}$&{\rm Top-Hat} \\&&\\\hline
&&\\$\gamma$&$1\quad {\rm or}\quad 2$ &{\rm Depends to the kinds of
Galaxy}\\&&\\ \hline
\end{tabular}
\end{center}

\begin{figure}
\centerline{\begin{tabular}{c}
\epsfig{figure=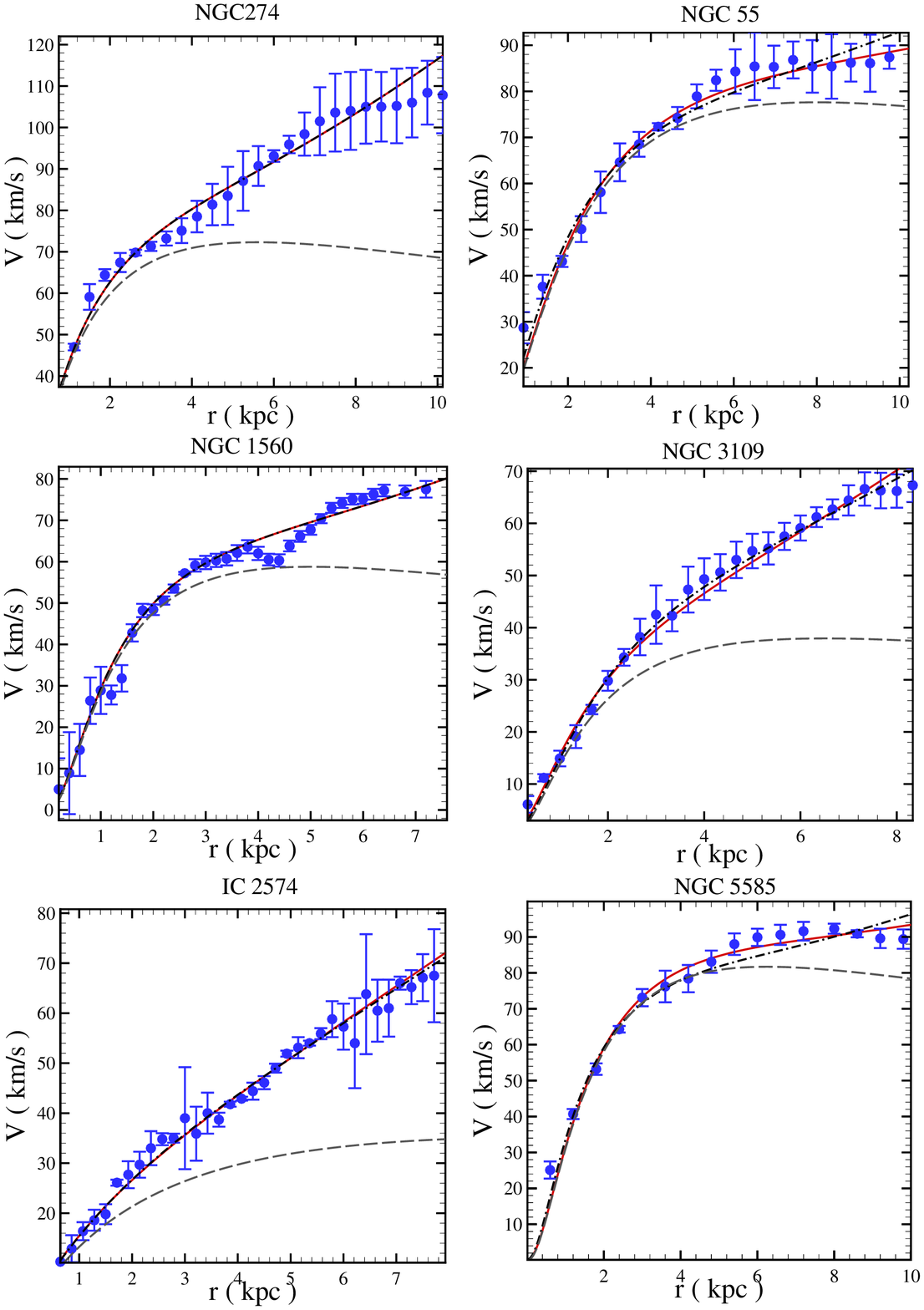,width=12.0cm}\\
\epsfig{figure=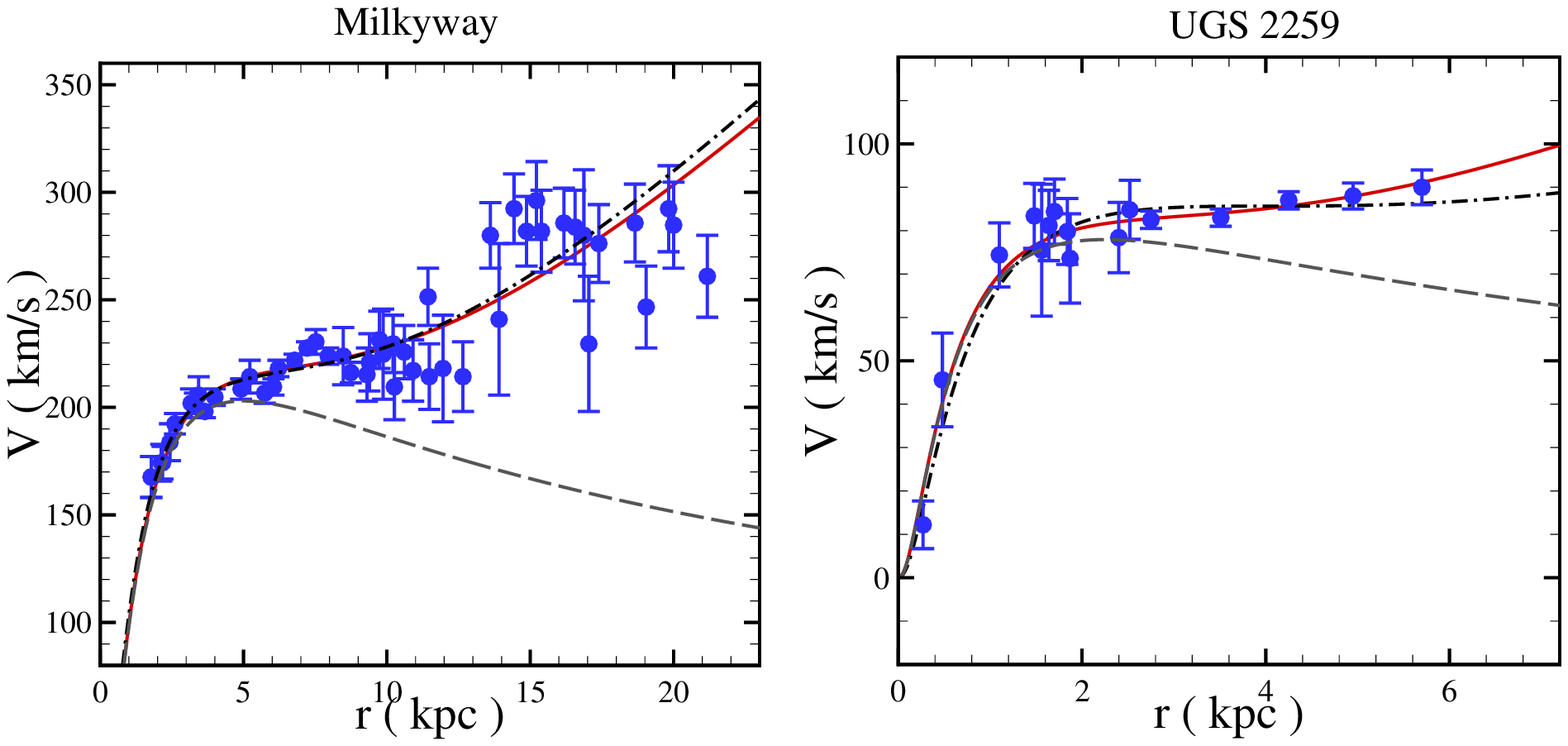,width=12.0cm}
\end{tabular}}
\caption{{\footnotesize The rotational velocity of various
galaxies in terms of distance. The solid line corresponds to the
theoretical prediction, equation (\ref{r67}), with the best fit
values for the free parameters derived by using Top-Hat prior for
$\bar{\beta}$. The theoretical fitting corresponding to the
Gaussian prior for $\bar{\beta}$ has been indicated by the
dashed-dot curves. The dashed curve shows the Newtonian-Keplerian
prediction. Data from observations indicated by symbols are for
the following galaxies: NGC 247, NGC 55, NGC 1560, NGC 3109, IC
2574, NGC 5585, UGC 2259 and Milkyway.}} \label{fig1}
\end{figure}

The best-fit values for the parameters of the model at
$1\sigma(68.3\%)$ and $2\sigma(95.4\%)$ confidence intervals with
corresponding reduced $\chi_{\nu}^2=\chi^2/N$ ($N$ is number of
degrees of freedom) for various galaxies are presented in Table
III. Table IV shows the best-fit values, using the Gaussian prior
for $\bar{\beta}$. To infer the Gaussian prior, we rely on the
determination of $\bar{\beta}$ from empirical measurement of the
time evolution of $G$ mentioned in section 4. Figure \ref{fig1}
compares the fitting rotational velocity curves derived by using
the best fitting parameters of some galaxies with observational
data points.

Since a proper theory should have a constant $\beta$, we infer a
mean value with its variance at $1\sigma$ and $2\sigma$ confidence
level for $\bar{\beta}$ from the best fit-values reported in Table
III and IV as
$$\bar{\beta}=6.38^{+2.44}_{-3.46}\!_{-6.71}^{+6.18}$$ and $$\bar{\beta}=5.72_{-1.18}^{+1.22}\!_{-2.69}^{+2.90}.$$
The fist value corresponds to the Top-Hat prior for $\bar{\beta}$
and the second is related to the Likelihood analysis with Gaussian
prior. As mentioned before, due to some systematic and random
errors in the observational data, one cannot expect the Likelihood
analysis to give a unique value for $\beta$ for our theoretical
model so that one is inclined to find a confidence interval. On
the other hand we should point out that in spite of many attempts
in previous models \cite{uzan}, we have found a mostly confined
interval for $\beta$ which is in agrement with pervious
predictions such as empirical measurement of the time evolution of
$G$. In addition, Weyl-Dirac theory gives a better prediction for
rotational velocity curves than Newtonian theory.

Note that the plots in figure 1 give the impression as if the
velocity curves tend to bend slightly upwards. However, this is
not the case since from the gravitational potential (\ref{r3})
obtained for an extended object such as a galaxy, we may infer
that there is a limiting radius for the formation of galaxies. In
fact, this gravitational potential differs from the Newtonian one
by two repulsive correction terms. For distances near the center
of a galaxy the effect of the first term is important. However at
larger distances the first term falls off and the potential is
dominated by the other two terms. There is a limiting distance $r$
for which we have $V(r)=0$ and far from it the force becomes
repulsive and the solar orbit becomes unstable. Therefore, we see
that the Weyl-Dirac theory predicts a limiting radius for a
galaxy.

The observational Tully-Fisher relation \cite{Tully} implies that
$v_{out}^{4}\sim G_{0}M$ where $v_{out}$ is the observed velocity
at the outermost observed radial position and $M$ is the galaxy
mass. Milgrom's phenomeological law \cite{milg3} or MOND model
predict the Tully-Fisher relation by assuming that the mass to
luminosity ratio, M/L, is constant across all galaxies. In
Milgrom's phenomenological model we have
\begin{equation}\label{mil}
 v^{4}_{out}=G M a_{0},
\end{equation}
at sufficiently low acceleration $a\ll a_{0}$ with $a_{0}\sim
10^{-8}$ $cm/s$. Relation (\ref{mil}) does depend on the magnitude
of the acceleration $a_{0}$, but not on the radial distance $r$
and the rotation velocity is constant out to an infinite range. In
contrast, our model does depend on the radius $r$ and the time
variation rate of the gravitational coupling. Despite the beauty
of Milgrom's law in explaining flat rotation curves of galaxies it
seems problematic to embed the theory within a comprehensive
relativistic theory of gravity. Therefore it is not clear that the
theory is as successful for explaining gravitational lensing of
clusters and other curved space time effects. In fact strong
gravitational lensing indicates a larger mass concentration at
cluster centers than accounted for by the present form of
Milgrom's theory \cite{sav}.

The properties of observational rotational velocity data points
for some famous galaxies used in our model are summarized in Table
II. In order to make a comparision of our results for galaxy
masses with other approaches we also list the mass obtained by
Milgrom phenomenological MOND model and the Moffat relativistic
MSTG model.
\begin{center}
\textbf{Table II:}\vspace{2mm}\\
\begin{tabular}{|c|c|c|c|c|}
\hline Galaxy & Surface  &
$M(10^{10}M_{\odot})$ & $M(10^{10}M_{\odot})$ & ref.\\
 &Brightness     &MOND&MSTG&\\

\hline NGC 1560& LSB \small{(Dwarf)}& 0.59$\pm$0.05&
0.79$\pm$0.05& \cite{br}\\

\hline NGC 3109 & LSB \small{(Dwarf)} &0.62$\pm$0.04 &
0.78$\pm$0.04&
\cite{jc}\\

\hline NGC 55   & LSB \small{(Dwarf)} &0.91$\pm$0.07&1.17$\pm$0.07&
\cite{pcw}\\

\hline UGC 2259 & HSB \small{(Dwarf)} &0.55$\pm$0.02&0.77$\pm$0.02 &
\cite{gsva}\\

\hline NGC 5585 & HSB \small{(Dwarf)}  & 0.9$\pm$0.06 &1.17$\pm$0.07
& \cite{cocs}\\

\hline NGC 247 & HSB &   1.46$\pm$0.14 & 2.27$\pm$0.17&
\cite{cp}\\

\hline IC 2574 & HSB &   -    & - & \cite{mcr}\\

\hline Milkyway & HSB & 10.60$\pm$0.37&9.12$\pm$0.28 &
\cite{mcr}\\

 \hline
\end{tabular}
\end{center}
\vspace{.5cm}\noindent\\

Column six shows the references used to extract the observational
data. Column three shows the values $M$ in $10^{10} M_{\odot}$
which are used to fit the velocity curves in our model. In columns
four and five we present the galaxy masses obtained from the MOND
model and the MSTG model \cite{mof}. Comparison of the results in
column 3 with that of columns 4 and 5 shows that the masses
obtained by the Weyl-Dirac theory are within the range obtained
from the MOND and MSTG models.

\section{Intergalactic effects}
The galaxy M31 and our own galaxy (the Galaxy) are dynamically the
only members of the so-called Local Group. Astronomical
observations show that the center of M31 and the Galaxy approach
each other with a speed of 119~m/s \cite{bt}. A possible
explanation of this phenomena was given by Kahn and Woltjer
\cite{kw}. They showed that the effect requires the reduce mass of
M31 and the Galaxy to be $100$ times larger. This discrepancy is
usually accounted for by the incorporation of dark matter.
Recently the effect has been explained by the nonsymmetric
gravitational theory \cite{ms}. Here we attempt to explain the
effect by using the Weyl-Dirac theory without resorting to the
dark matter hypothesis.

We first note that in the absence of intergalactic matter, M31 and
the Galaxy form  a double galaxy. In this case, using (\ref{r3}),
the equation of motion for the system becomes
\begin{equation}\label{E1}
\ddot{r}+
\frac{M_{LG}G}{r^{2}}+\frac{M_{LG}\beta^{2}}{M_{\odot}}r=0\,
\end{equation}
where $M_{LG}$ is the total mass of M31 and the Galaxy and $r$ is
the relative distance between them. Because the distance between
M31 and the Galaxy is about 700 kpc, the term
$\frac{M_{LG}G}{r^{2}}$ in (\ref{E1}) is negligible and can be
ignored. In this case we have

\begin{equation}\label{E2}
\ddot{r}+\frac{M_{LG}\beta^{2}}{M_{\odot}}r=0\,.
\end{equation}
This equation is equivalent to the equation of motion derived by
Kahn and Woltjer by assuming the intergalactic dark matter
spreading homogeneously through the Local Group with mass density
$\rho\equiv \frac{3M_{LG}}{4\pi G M_{\odot}} \beta^{2}$.
Therefore, it appears that the Weyl-Dirac theory may help us to
make a better understanding of the intergalactic effects without
the exotic dark matter hypothesis.

\section{Gravitational lensing}

The gravitational lensing or deflection of light by massive bodies
provides another way to test the relativistic effects of
gravitational theories at large scales. The gravitational lensing
gives the most accurate information about the mass of the core of
galaxies \cite{tvw} and the cosmic telescope effects of
gravitational lenses have enabled us to study faint and distant
galaxies which happen to be strongly magnified by galaxy clusters.
Also, the statistics of gravitational lensing events can be used
to estimate cosmological parameters \cite{gn}. Observations of
strong lensing by clusters indicate a larger mass concentration at
cluster centers than their visible matter \cite{bn}. This
discrepancy is usually accounted for by the existence of dark
matter. Here we describe this effect by the Weyl-Dirac theory.

The deflection angle of light rays passing through a gravitational
potential $V$ is given by

\begin{equation}\label{g1}
\alpha=\frac{2}{c^{2}}\int_{s}^{o} \nabla_{\bot} V dl,
\end{equation}
where the integral is evaluated along the path traversed by light
joining the source (s) to distant observer (o) and $\nabla_{\bot}$
denotes the derivative in the direction perpendicular to the path.
Let us now consider deflection the of light from a cluster of
galaxies. At these scales we can safely use the point mass
approximation (\ref{r2}). Clearly, to study microlensing effects
for smaller scales, we may use gravitational potential (\ref{r3}).
Using (\ref{r2}), we find the deflection angle

\begin{equation}\label{g2}
\alpha= \frac{4GM}{c^{2}b}\left(1+\frac{b^{2}\beta^{2}d_{os}}{2M
G}\right),
\end{equation}
where $b$ and $d_{os}$ denote the impact parameter and the
distance between the observer and source, respectively. Relation
(\ref{g2}) shows a larger light deflection than is expected from
standard gravitational theories. For example for a cluster with
$b\sim 4\times 10^{4}$ kpc ,  $d_{os}=10$ Gpc and $\beta\sim
0.74\times 10^{-13}$ /yr we obtain

\begin{equation}\label{g3}
\frac{b^{2}\beta^{2}d_{os}}{2M_{\odot}G}\sim 10\,.
\end{equation}
Therefore, the increase in the light deflection may be explained
without the dark matter hypothesis.

\section{Conclusions}

We have considered the Weyl-Dirac theory and showed that the
spherically symmetric solutions of the theory lead to a
gravitational potential which differs from the Newtonian one due
to the appearance of a repulsive correction term increasing with
distance. The correction term appears to be determined by the time
variation rate of the gravitational coupling. We saw that the
correction term for an extended massive source such as a galaxy
becomes more important, especially at galactic distances. Hence
one may expect the theory to imply significant consequences for
galaxies and on galactic scales.

We also performed well-known Likelihood analysis on the free
parameters in our model to check the consistency of our
predictions to that of the observational data. In doing so, we
considered eight galaxies which are chosen from different types,
namely the low and the high surface brightness galaxies. Using two
different types of prior for model parameters, the best fit-values
for free parameters at $1\sigma(68.3\%)$ and $2\sigma(95.4\%)$
confidence intervals have been determined. The mean value for
$\bar{\beta}$ using Top-Hat and Gaussian priors are
$6.38^{+2.44}_{-3.46}\!_{-6.71}^{+6.18}$ and
$5.72_{-1.18}^{+1.22}\!_{-2.69}^{+2.90}$, respectively. However we
found a relatively wide  interval for the best fit-values of
$\bar{\beta}$ for each galaxy, but this gives a small deviation in
the goodness of the fitting curves with observation, in other
words the goodness of the fit is almost insensitive to major
fluctuations of $\bar{\beta}$. Also Likelihood analysis lets us to
obtain a much more smaller confidence interval for $\bar{\beta}$
than the former extracted values which are also consistent with
previous inferring. The fit to data seems to be acceptable in
terms of the total mass $M$ of each galaxy and a best fit-value
for $\beta$ without the exotic dark matter hypothesis at the
expense of a rather wide interval range for $\beta$. This could be
regarded as the main disadvantage of the model presented here.
However, it should also be pointed out that the goodness of the
fit is acceptable for the underlaying observational data set used
in this analysis.

\begin{table}
\begin{center}
\textbf{Table III:  Best fitting parameters at $1\sigma$ and
$2\sigma$ confidence levels with corresponding $\chi_{\nu}^2$ for
various
galaxies.}\vspace{2mm}\\
\begin{tabular}{|c|c|c|c|c|c|}
\hline {\rm Galaxy} & $\bar{\alpha}$&$\bar{\beta}$&$r_c$\quad{\rm
kPc}&$\gamma$&$\chi_{\nu}^2$
\\\hline
&&&&&\\ {\rm NGC 1560}
&$1.20_{-0.20}^{+0.05}$&$6.95_{-0.10}^{+1.25}$
&$1.00_{-0.10}^{+0.05}$&$2$&$3.26$\\&&&&&\\
&$1.20_{-0.35}^{+0.10}$&$6.95_{-0.55}^{+2.15}$
&$1.00_{-0.20}^{+0.10}$&&\\&&&&&
\\\hline

&&&&&\\ {\rm NGC 3109}
&$0.65_{-0.10}^{+0.25}$&$9.40_{-2.10}^{+1.00}$
&$1.30_{-0.10}^{+0.15}$&$2$&$0.99$\\&&&&&\\
&$0.65_{-0.25}^{+0.50}$&$9.40_{-3.50}^{+4.10}$
&$1.30_{-0.20}^{+0.30}$&&\\&&&&&
\\\hline

&&&&&\\ {\rm NGC 55} &$3.35_{-0.30}^{+0.30}$&$2.50_{-0.55}^{+0.50}$
&$1.60_{-0.05}^{+0.05}$&$2$&$0.86$\\&&&&&\\
&$3.35_{-0.65}^{+0.80}$&$2.50_{-1.30}^{+1.00}$
&$1.60_{-0.15}^{+0.15}$&&\\&&&&&
\\\hline

&&&&&\\ {\rm UGC 2259}
&$0.95_{-0.10}^{+0.15}$&$11.30_{-2.30}^{+0.55}$
&$0.45_{-0.05}^{+0.05}$&$2$&$0.46$\\&&&&&\\
&$0.95_{-0.20}^{+0.25}$&$11.30_{-4.60}^{+1.10}$
&$0.45_{-0.10}^{+0.15}$&&\\&&&&&
\\\hline

&&&&&\\ {\rm NGC 5585}
&$2.90_{-0.30}^{+0.25}$&$3.05_{-0.20}^{+0.70}$
&$1.25_{-0.10}^{+0.05}$&$2$&$2.39$\\&&&&&\\
&$2.90_{-0.55}^{+0.55}$&$3.05_{-0.80}^{+1.25}$
&$1.25_{-0.15}^{+0.10}$&&\\&&&&&
\\\hline

&&&&&\\ {\rm NGC 247} &$2.30_{-0.25}^{+0.25}$&$6.35_{-0.40}^{+0.50}$
&$2.80_{-0.15}^{+0.15}$&$1$&$1.05$\\&&&&&\\
&$2.30_{-0.45}^{+0.35}$&$6.35_{-0.90}^{+1.40}$
&$2.80_{-0.30}^{+0.30}$&&\\&&&&&
\\\hline

&&&&&\\ {\rm IC  2574}
&$1.05_{-0.25}^{+0.45}$&$7.95_{-1.50}^{+1.50}$
&$5.35_{-0.40}^{+0.65}$&$1$&$1.63$\\&&&&&\\
&$1.05_{-0.45}^{+1.00}$&$7.95_{-2.85}^{+3.30}$
&$5.35_{-0.95}^{+1.50}$&&\\&&&&&
\\\hline

&&&&&\\ {\rm Milkyway}
&$14.32_{-0.15}^{+0.20}$&$3.55_{-0.15}^{+0.10}$
&$1.00_{-0.05}^{+0.05}$&$2$&$1.75$\\&&&&&\\
&$14.32_{-1.05}^{+1.20}$&$3.55_{-0.25}^{+0.35}$
&$1.00_{-0.10}^{+0.10}$&&\\&&&&&
\\\hline
\end{tabular}
\end{center}
\end{table}

\begin{table}
\begin{center}
\textbf{Table IV:  Best fitting parameters at $1\sigma$ and
$2\sigma$ confidence levels with corresponding $\chi_{\nu}^2$ for
various
galaxies using the Gaussian prior for $\bar{\beta}$ in the Likelihood analysis.}\vspace{2mm}\\
\begin{tabular}{|c|c|c|c|c|c|}
\hline {\rm Galaxy} & $\bar{\alpha}$&$\bar{\beta}$&$r_c$\quad{\rm
kPc}&$\gamma$&$\chi_{\nu}^2$
\\\hline
&&&&&\\ {\rm NGC 1560}
&$1.20_{-0.10}^{+0.10}$&$6.95_{-0.10}^{+0.10}$
&$1.00_{-0.05}^{+0.05}$&$2$&$3.29$\\&&&&&\\
&$1.20_{-0.15}^{+0.15}$&$6.95_{-0.55}^{+0.65}$
&$1.00_{-0.10}^{+0.10}$&&\\&&&&&
\\\hline

&&&&&\\ {\rm NGC 3109}
&$1.00_{-0.10}^{+0.10}$&$6.75_{-0.45}^{+0.35}$
&$1.50_{-0.10}^{+0.05}$&$2$&$1.09$\\&&&&&\\
&$1.00_{-0.20}^{+0.25}$&$6.75_{-1.05}^{+1.35}$
&$1.50_{-0.15}^{+0.20}$&&\\&&&&&
\\\hline

&&&&&\\ {\rm NGC 55} &$2.45_{-0.15}^{+0.25}$&$4.05_{-0.55}^{+0.25}$
&$1.40_{-0.05}^{+0.10}$&$2$&$1.93$\\&&&&&\\
&$2.45_{-0.35}^{+0.60}$&$4.05_{-0.95}^{+0.70}$
&$1.40_{-0.10}^{+0.15}$&&\\&&&&&
\\\hline

&&&&&\\ {\rm UGC 2259}
&$1.30_{-0.05}^{+0.10}$&$6.70_{-0.50}^{+0.80}$
&$0.55_{-0.05}^{+0.05}$&$2$&$0.79$\\&&&&&\\
&$1.30_{-0.15}^{+0.20}$&$6.70_{-1.30}^{+1.45}$
&$0.55_{-0.10}^{+0.10}$&&\\&&&&&
\\\hline

&&&&&\\ {\rm NGC 5585}
&$2.15_{-0.05}^{+0.20}$&$4.60_{-0.20}^{+0.15}$
&$1.10_{-0.05}^{+0.05}$&$2$&$3.29$\\&&&&&\\
&$2.15_{-0.25}^{+0.45}$&$4.60_{-1.00}^{+0.70}$
&$1.10_{-0.10}^{+0.10}$&&\\&&&&&
\\\hline

&&&&&\\ {\rm NGC 247} &$2.30_{-0.15}^{+0.15}$&$6.35_{-0.40}^{+0.45}$
&$2.80_{-0.15}^{+0.10}$&$1$&$1.05$\\&&&&&\\
&$2.30_{-0.35}^{+0.35}$&$6.35_{-0.85}^{+1.05}$
&$2.80_{-0.25}^{+0.25}$&&\\&&&&&
\\\hline

&&&&&\\ {\rm IC  2574}
&$1.40_{-0.20}^{+0.20}$&$6.60_{-0.55}^{+0.65}$
&$5.95_{-0.65}^{+0.70}$&$1$&$1.65$\\&&&&&\\
&$1.40_{-0.35}^{+0.35}$&$6.60_{-1.15}^{+1.39}$
&$5.95_{-0.65}^{+0.70}$&&\\&&&&&
\\\hline

&&&&&\\ {\rm Milkyway}
&$13.45_{-0.15}^{+1.05}$&$3.80_{-0.35}^{+0.10}$
&$0.95_{-0.05}^{+0.05}$&$2$&$2.09$\\&&&&&\\
&$13.45_{-0.80}^{+1.30}$&$3.80_{-0.45}^{+0.20}$
&$0.95_{-0.10}^{+0.10}$&&\\&&&&&
\\\hline
\end{tabular}
\end{center}
\end{table}

\end{document}